\documentclass[conference,a4paper]{IEEEtran}
\IEEEoverridecommandlockouts
\usepackage{amsmath,amssymb,amsfonts}
\usepackage{bbm} 
\usepackage{algorithmic}
\usepackage{graphicx}
\usepackage{textcomp}
\usepackage{xcolor}
\usepackage[style=ieee]{biblatex}
\def\BibTeX{{\rm B\kern-.05em{\sc i\kern-.025em b}\kern-.08em
    T\kern-.1667em\lower.7ex\hbox{E}\kern-.125emX}}
 
\begin{document}

\title{Multilevel Monte Carlo with Surrogate Models for Resource Adequacy Assessment\\
}

\author{\IEEEauthorblockN{Ensieh~Sharifnia}
\IEEEauthorblockA{\textit{Dept. of Electrical Sustainable Energy} \\
\textit{Delft University of Technology}\\
Delft, The Netherlands \\
e.sharifnia@tudelft.nl}
\and
\IEEEauthorblockN{Simon~H.~Tindemans}
\IEEEauthorblockA{\textit{Dept. of Electrical Sustainable Energy} \\
\textit{Delft University of Technology}\\
Delft, The Netherlands \\
s.h.tindemans@tudelft.nl}
}

\IEEEoverridecommandlockouts

\IEEEpubid{\parbox{\columnwidth}{\copyright 2022 IEEE. Personal use of this material is permitted. Permission from IEEE must be obtained for all other uses, in any current or future media, including reprinting/republishing this material for advertising or promotional purposes, creating new collective works, for resale or redistribution to servers or lists, or reuse of any copyrighted component of this work in other works.} \hspace{\columnsep}\makebox[\columnwidth]{ }}

\maketitle

\allowdisplaybreaks

\IEEEpubidadjcol

\begin{abstract}
Monte Carlo simulation is often used for the reliability assessment of power systems, but it converges slowly when the system is complex. Multilevel Monte Carlo (MLMC) can be applied to speed up computation without compromises on model complexity and accuracy that are limiting real-world effectiveness. In MLMC, models with different complexity and speed are combined, and having access to fast approximate models is essential for achieving high speedups. This paper demonstrates how machine-learned surrogate models are able to fulfil this role without excessive manual tuning of models. Different strategies for constructing and training surrogate models are discussed. A resource adequacy case study based on the Great Britain system with storage units is used to demonstrate the effectiveness of the proposed approach, and the sensitivity to surrogate model accuracy. The high accuracy and inference speed of machine-learned surrogates result in very large speedups, compared to using  MLMC with hand-built models.  

\end{abstract}


\section{Introduction}

Probabilistic evaluation techniques are utilized for reliability assessment in the power system because they reflect the stochastic nature of the power system and its variables~\cite{li2013reliability}. Among evaluation techniques for large and complex systems, Monte Carlo (MC) methods are very flexible. MC methods estimate the risk indices by simulating the actual process and considering random behavior of the system,  but may converge slowly when estimating rare event risks~\cite{li2013reliability, zhao2021cross}. This is a concern, especially in highly reliable power systems.
Variance reduction techniques, which decrease the variance of estimation, are useful in these applications~\cite{li2013reliability, billinton1996variance}. Multilevel Monte Carlo (MLMC) is one of the powerful variance reduction techniques that has recently been used in reliability context~\cite{aslett2017multilevel, huda2017improving}. The MLMC approach can be used for resource adequacy studies based on either snapshot or time-sequential simulations ~\cite{tindemans2020accelerating}.

Multilevel Monte Carlo reduces the variance of estimation and improves computational efficiency of simulation by applying models with different complexity. Higher speedups can be achieved when models in level pairs have high correlation and each simplified model is much faster than the next level~\cite{giles2015multilevel}. Therefore, having good combinations of models has a significant effect on MLMC performance. However, manually constructing models with the above conditions~\cite{tindemans2020accelerating} may require substantial domain knowledge.

Surrogate models have been applied to approximate results of complex simulations and to reduce the computational barrier in simulation-based design optimization~\cite{li2019surrogate}. Prediction accuracy varies over the input domain and highly depends on the size of training data. A major concern when using surrogate modeling - especially for reliability assessment of power systems - is quantifying and propagating the model's uncertainty, which may result in less confidence in the prediction of failure probability~\cite{li2019surrogate}. 

This paper proposes to use MLMC with surrogate models to combine their benefits and overcome their respective limitations: surrogate models provide fast and highly correlated `black box' models and MLMC delivers unbiased estimation with confidence intervals. Different approaches for designing and training surrogate models for adequacy assessment are discussed. A variety of MLMC model architectures is compared in a case study, demonstrating the flexibility of the approach and the ability to attain very large speedups. Finally, the impact of training size on surrogate accuracy and overall speedup is investigated.

\section{MLMC for adequacy risk assessment}

\subsection{Risk measures}
Power system adequacy is usually quantified using risk measures in the form of expectation values, i.e. $q = E[X(Z)]$. Here, $X$ is a performance function that quantifies the degree of (non-)performance of states $Z$ from a sample space $\Omega$ and associates a real number with each state ($X:\Omega \rightarrow \mathbb{R}$). Monte Carlo (MC) simulation methods estimate the risk index $q$ by randomly sampling states $Z$ from $\Omega$ according to the probabilistic model of the system and evaluating them based on the performance function $X$.
Loss of Load Expectation (LOLE) and Expected Energy Not Served (EENS) are two commonly used resource adequacy risk measures. They are defined as follows, based on annual traces of load curtailment $c_t(Z)$ in hour $t$: 
\begin{align}
    \label{eq:LOLE}
    \mathrm{LOLE} &= E\left[ \sum_{t=1}^{8760}\mathbbm{1}_{c_t(Z)} \right], \\
    \label{eq:EENS}
    \mathrm{EENS} &= E\left[\sum_{t=1}^{8760}{c_t(Z) \times 1 h}\right].
\end{align}
\subsection{Multilevel Monte Carlo}

Risk measures of the form $q = E[X(Z)]$ can be efficiently estimated using MLMC~\cite{giles2015multilevel, tindemans2020accelerating}. The framework from~\cite{tindemans2020accelerating} is summarised in this section, and we note that the mathematical analysis can be extended for estimating multiple risk indices $q_a, q_b, ...$ (e.g. LOLE, EENS), through using different performance functions $X_{(a)}, X_{(b)}, ...$ in parallel. 

In the remainder of the paper, we will omit the explicit dependence on the random state $Z$ and directly analyse the random variables $X\equiv X(Z)$, i.e. performance metrics of the system. In the following, random variables are denoted using upper case characters and scalars and deterministic values are indicated using lower case characters.

MLMC combines samples from models of different complexity to achieve speed-ups without compromising accuracy. Consider a set of $L$ models $M_1, M_2, ..., M_L$ of the \emph{same} system, which increase in complexity and generate output variables $X_1, ..., X_L$, respectively. The expectation of the top level model $q=E[X_L]$ is the quantity of interest, but evaluating this model is computationally demanding. The lower level models $M_1, \ldots, M_{L-1}$ generate output variables $X_1, ..., X_{L-1}$ that are increasingly accurate approximations of $X_L$. MLMC uses these approximate models to better estimate $E[X_L]$. We have
\begin{align}
\label{eq:telescopic}
    q &= E[X_L] \nonumber \\
      &= E[X_1] + E[X_2 - X_1] + ... +E[X_L - X_{L-1}] \nonumber \\
      &= r_1 + ... + r_L,
\end{align}
where $r_l$ is the contribution for level $l$. $r_1$ can be considered a crude estimation of $q$ and $r_2, ..., r_L$ are successive refinements. 

In MLMC, each level contribution $r_l$ is estimated independently by means of MC simulation:
\begin{align} \label{eq:levelcontribution}
    \hat{R}_l &= {\frac{1}{n_l}}\sum_{i=1}^{n_l}{Y_l^{(i)}},
\intertext{with}
    Y_l^{(i)} &= X_l^{(l,i)} - X_{l-1}^{(l,i)},\\
    X_{0} &\equiv 0.
\end{align}
Here, $X_l^{(k,i)}$ are independently sampled outputs from model $l$. We note that the superscript $k\in\{l, l+1 \}$ is added to the level outcome $X_l$ to indicate that the outputs for model $l$ can be produced differently depending on whether they are paired with the output of higher model or lower model, as long as the expectations are equal: $E[X_l] = E[X_l^{(l+1,i)}] = E[X_l^{(l,i)}]$. Combining \eqref{eq:telescopic} and \eqref{eq:levelcontribution}, the MLMC estimator  $\hat{Q}$ is defined as
\begin{align}
\label{eq:Q_MLMC}
    \hat{Q} \equiv \sum_{l=1}^{L}\hat{R}_l  = \sum_{l=1}^L {\frac{1}{n_l}}\sum_{i=1}^{n_l}{Y_l^{(i)}}.    
\end{align}

In MLMC, samples for estimating each $\hat{R}_l$ are randomly and independently selected; only samples in level pairs $X_l^{(l,i)}, X_{l-1}^{(l,i)}$ are jointly selected from a common distribution. Invoking the central limit theorem for each level pair, we know that the MLMC estimator \eqref{eq:Q_MLMC} is unbiased and asymptotically normally distributed with variance
\begin{align}
\label{eq:var_MLMC}
    \sigma_{\hat{Q}}^2 &= \sum_{l=1}^{L} \frac{\sigma_{Y_l}^2}{n_l}, \quad \textrm{where}\\
\label{eq:var_level}
    \sigma_{Y_l}^2 &= \sigma_{X_l^{(l)}}^2 +\sigma_{X_{l-1}^{(l)}}^2 -2 \mathrm{cov}(X_l^{(l)},X_{l-1}^{(l)} ).
\end{align}

According to (\ref{eq:var_level}), the variance of estimation decreases when samples correlation in level pairs increases. The performance of MLMC is also dependent on the number of samples $n_l$ that are generated in each level. The optimal number of samples $n_l^*$ for each level pair is given by~\cite{tindemans2020accelerating}
\begin{align}
\label{eq:n*}
n_l^* = \frac{t }{\sum_{l'=1}^L {\sigma_{Y_{l'}}}\sqrt{\tau_{l'}}}\times \frac{\sigma_{Y_l}}{\sqrt{\tau_l}},
\end{align}
where $t$ is the computation budget and $\tau_l$ is the approximated time for generating a sample realization $y_l$.

It is worth mentioning that the MLMC method reduces to the conventional MC method when only one layer (the top layer) is used. In this way, all samples are selected from the top layer and there is no rough estimation and refinement. 

\subsection{Speed measure}
The asymptotic efficiency of different sampling-based estimators of $q$ (including different MLMC level stacks) can be compared with the speed measure~\cite{tindemans2020accelerating}
\begin{align}
    \label{eq:speed}
    z_q = \frac{q^2}{t \sigma_{\hat{Q}}^2},
\end{align}
where $t$ is the execution time of simulation. The main advantage of using (\ref{eq:speed}) is that the speeds corresponding to different risk measures are directly comparable when a given coefficient of variation is targeted. Therefore, it indicates which risk measure is the limiting factor for the simulation to achieve a desired level of accuracy. Moreover, the ratio of speeds gives the asymptotic speedup of one method over another.

\section{MLMC with learned surrogate models}

Instead of simulation techniques, surrogate models have been applied to provide fast and close approximation of performance function's outcome~\cite{li2019surrogate}. However, because the approximation errors (and particular bias) of surrogate models are often unknown, using surrogate models to directly estimate risk may not be advisable. 

MLMC is a general framework that can be used to achieve bias-free estimates of risk measures with arbitrary approximate models. A high speedup can be achieved, when level pairs ($X_l,X_{l-1}$) have high correlation and each simplified model is much faster than the next level up~\cite{tindemans2020accelerating}. However, defining models with low computational overhead that closely approximate the model of interest can be a labour-intensive process. Hence, we propose to use MLMC with surrogate models, benefiting from fast evaluation and close approximation of surrogate models without laborious modelling.

\subsection{Proposed approach}

There are two distinct approaches in which surrogate models can be applied in MLMC methods for risk assessment. One approach, alluded to above, is to use them independently as a level in MLMC to approximate a performance function. This way, a large speedup can be achieved without the need to define bottom-up approximate models. The second approach is to integrate surrogate models with hand-tuned models in a way that the produced model can execute faster. Consequently, higher speedups with MLMC can be obtained. In this paper, we apply both approaches for resource adequacy study.

A consideration when using surrogate models is the choice of the feature space on which the model is trained. This may be straightforward for snapshot simulation, where the variables defining the current network state can be treated as a sample and the surrogate model approximates the corresponding performance function. It is less easy for time-sequential simulation, where the trajectory of the state affects the next state and the whole sequence determines the outcome of the performance function. It is usually not efficient (or impossible) to use the whole sequence of states as an input for the surrogate model when estimating the outcome of the performance function.  

To apply surrogate models for sequential simulation, we propose to train a surrogate model on short sequences of \emph{exogenous} variables. The proper size for the frame has two important features. 1) It is the smallest size that results in the minimum state dependency for the subsequent time frame. 2) The state sequence within the time frame usually follows a similar pattern. Choosing the proper size for the time frame is dependent on the application. For example, in resource adequacy assessment a suitable time frame may be 24 hours because usually load curtailment does not occur during midnight (the first feature) and most days have similar patterns (the second feature).

\subsection{Training strategies}

Surrogate models are data-driven and therefore require a training strategy. There are two approaches for training a model: 1) one-off training, and 2) adaptive learning. For one-off training, training data is first selected from the state space $\omega$ (heuristically or randomly). Second, the performance function is computed based on the detailed model. Finally, the surrogate model is trained and used as a level within MLMC estimation. Adaptive learning also requires an initial training phase, but with a relatively small number of training samples. After this short training phase, the model is used in MLMC and is tuned during the simulation by those samples that are generated from the top level. In this paper, the former approach is used.

\begin{figure*}[t!b]
\centerline{\includegraphics{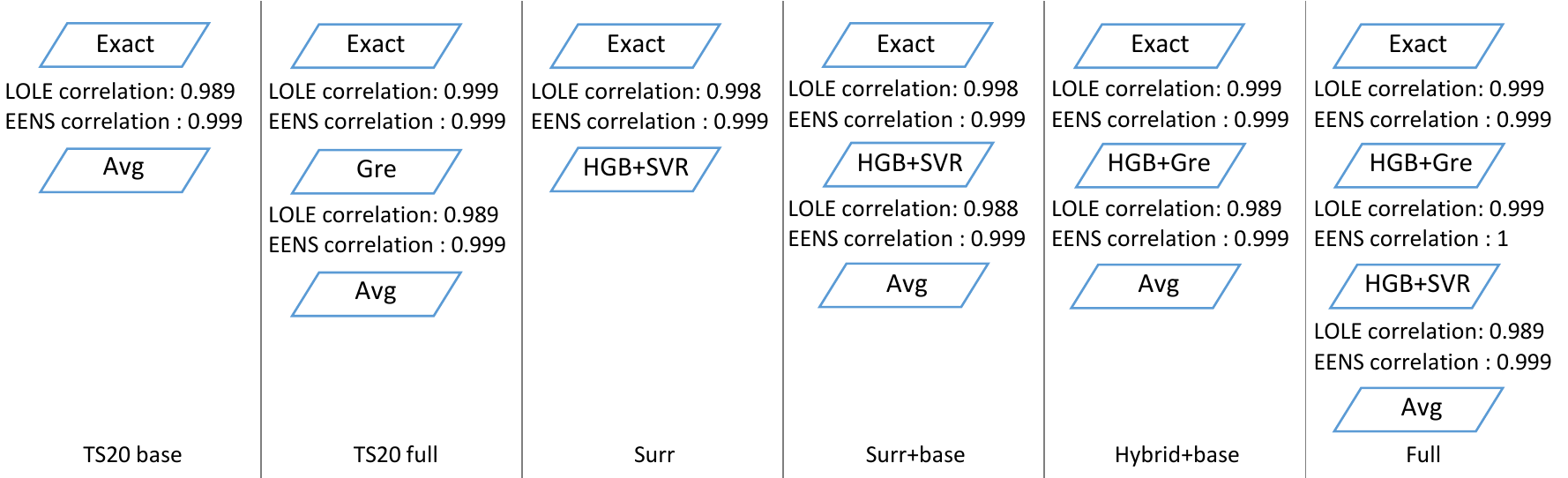}}
\caption{MLMC model combinations used. Level-to-level sample correlations for LOLE and EENS are indicated.}
\label{fig:MLMC_correlation}
\end{figure*}

\section{Case study: generation adequacy with storage}
To demonstrate the effectiveness of using surrogate models with MLMC, a resource adequacy case study involving storage dispatch is chosen. Analysing this model necessitates time-consuming sequential simulations, and is therefore a good candidate for efficiency improvements. We consider five dispatch policies which have the same sample space $\Omega$ (net generation margin traces, in this case), but are different in terms of accuracy and computational complexity. The random net generation margin traces are generated from uncertainty in wind and demand patterns and generator outages.

\subsection{System description} \label{sec:model}

The resource adequacy model used in this paper is based on data from the Great Britain (GB) system. It is summarised here, and we refer readers to~\cite{tindemans2020accelerating} for further details. Annual demand traces and wind traces are randomly drawn from historical data and a synthetic data set respectively. For the latter, we assumed a 10~GW wind generation capacity and constant distribution of wind generation sites. Conventional generation traces were generated by diverse thermal units. The portfolio of 27 storage units was derived from the T-4 GB capacity auction contract in 2018. For each sample year $i$, a demand trace $d_t^{(i)}$, wind power trace $w_t^{(i)}$ and conventional generation trace $g_t^{(i)}$ are drawn. They are combined into a net generation margin trace $m_t^{(i)}$ of 8760 hours (365 days):
\begin{align}
    \label{eq:net_margin}
    m_t^{(i)} = g_t^{(i)} + w_t^{(i)} - d_t^{(i)}, \qquad t \in \{1, ..., 8760\}.
\end{align}

This study considers the generation adequacy in the presence of battery storage. Modeling storage dispatch policy is a time-coupled problem, so chronological simulation is needed. The load curtailment $c_{t}^{(i)}$ in a simulated year $i$ is driven by the net generation margin trace $m_t^{(i)}$ and the resulting storage dispatch  $s_{t}^{(i)}\equiv s(m_{t}^{(i)})$:
\begin{align}
    \label{eq:load_curtailment}
    c_{t}^{(i)} = \max[0, -m_t^{(i)} + s_{t}^{(i)}], \qquad t \in \{1, ..., 8760\}
\end{align}

The reference dispatch policy is the EENS-minimising dispatch policy given in~\cite{evans2019minimizing}. The load curtailment following from this reference policy is included as the top layer for all MLMC simulations. In the following, this is referred to as the Exact model. 

\subsection{Simplified models}

Two simplified models are used to model the storage dispatch. They were designed to approximate the Exact reference policy at significantly reduced complexity~\cite{tindemans2020accelerating}. 

\subsubsection{Greedy dispatch}

In the Greedy dispatch (Gre) model, storage units are arranged in descending order by time-to-go from large (high energy-to-power ratio) to small. Then, in a greedy sequential pass per storage unit, batteries charge when possible, and discharge to avoid load curtailment.

\subsubsection{Average dispatch}

The Average dispatch (Avg) model makes two assumptions that simplify the storage dispatch model into a deterministic load offset, which means that efficient convolution operations can be used for estimating risk indices. The first assumption is considering all units as one big storage unit. The energy $\Bar{e}$ and the discharge power rate $\Bar{p}$ of this storage are equal to the summation of energies and discharge power rates of all units, respectively. The second assumption is considering the mean daily demand profile $\tilde{d}_{1:24}$ as a nominal demand profile. A repetitive 24h dispatch pattern is determined as a peak shaving solution by solving a quadratic optimization problem~\cite{tindemans2020accelerating}.

\subsection{Surrogate models}

In addition, two data-driven surrogate models are defined to estimate the adequacy performance of the system when operated with the Exact reference policy. To get an efficient feature representation, we make use of the fact that storage is nearly always able to recover fully overnight. Moreover, days have a similar pattern for net margin traces, and the margin tends to be positive at midnight.  

Hence, we expect that we can efficiently learn surrogate models for Loss Of Load (LOL) and Energy Not Supplied (ENS) by using daily margin traces as features. Surrogate models for LOL and ENS of an annual margin trace $m_{1:8760}^{(i)}$ are therefore constructed as 
\begin{align}
    \label{eq:Surr_LOL}
    f_{LOL}(m_{1:8760}^{(i)}) = \sum_{d=1}^{365} \tilde{f}_{LOL}(\tilde{m}_d^{(i)})
\end{align}
\begin{align}
    \label{eq:Surr_ENS}
    f_{ENS}(m_{1:8760}^{(i)}) = \sum_{d=1}^{365} \tilde{f}_{ENS}(\tilde{m}_d^{(i)}).
\end{align}
where $\tilde{m}_d^{(i)}$ represents the margin values for day $d$ from $m_{1:8760}^{(i)}$ and $\tilde{f}_{LOL}$ and $\tilde{f}_{ENS}$ are learned models. The models are trained on daily curtailment sequences $c_t^{(i)}$ that are obtained using the exact model, assuming that storage units were full at the start of the day.

\subsubsection{HGB+Greedy dispatch}

The HGB+Greedy (HGB+Gre) model consists of two parts. First, it applies the Histogram-based Gradient Boosting Regression Tree (HGBRT) estimator to predict the LOL of a given day (i.e. train $\tilde{f}_{LOL}$). HGBRT is an ensemble learner that sequentially adds tree models to the learner to correct the prediction errors of the learner like other boosting algorithms. HGBRT bins values of continuous features into a fixed number of buckets and uses them to construct feature histograms.  We choose the HGBRT algorithm because it approximates outcome of $\tilde{f}_{LOL}$ with high accuracy using a limited number of training samples.

For ENS prediction, this model hybridises HGBRT with the sequential greedy dispatch policy (Gre). First, HGBRT is used to estimate LOL, and only for those days where HGBRT predicts load curtailment events, the greedy dispatch is called to estimate the amount of ENS. Otherwise, the ENS is estimated at 0~MWh.

\subsubsection{HGB+SVR dispatch}

The HGB+SVR model is completely based on data-driven surrogate models. HGBRT is used to approximate $\tilde{f}_{LOL}$. For those days on which HGBRT predicts load curtailment events, Support Vector Regression (SVR) is called to estimate the amount of energy not served ($\tilde{f}_{ENS}$). Otherwise, the ENS is estimated at 0~MWh. The hyperplane construction approach used in SVR makes it robust against outliers, so it is a good choice for ENS values, which can vary in a wide range.

\begin{table*}[h!tb]
    \caption{Computational efficiency of MLMC model combinations}
    \renewcommand{\arraystretch}{1.5}
    \begin{tabular}{p{1.5cm} p{4.3 cm} p{1cm} p{2cm} p{2cm} p{1.9cm} p{1.9cm}}
    \hline
        Estimator & Architecture&Time (s)&LOLE (h/y)& EENS (MWh/y)&LOLE Speedup& EENS Speedup\\ \hline
        MC& Exact &26600&$1.745\pm0.039$&$2420.0\pm72.0$&n/a&n/a\\
        TS20 base &Exact $\mid$ Avg &26200&$1.743\pm0.009$&$2406.0\pm3.3$&18&482\\
        Surr & Exact $\mid$ HGB+SVR &25900&$1.739\pm0.005$&$2407.0\pm10.0$&47&49\\
        TS20 full &Exact $\mid$ Gre $\mid$ Avg &26000&$1.735\pm0.003$&$2405.0\pm1.2$&125&3617\\
        Surr$+$base & Exact $\mid$ HGB+SVR $\mid$ Avg &25900&$1.738\pm0.002$&$2408.0\pm1.1$&269&4775\\
        Hybrid$+$base & Exact $\mid$ HGB+Gre $\mid$ Avg &25800&$1.738\pm0.002$&$2405.0\pm0.5$&582&20946\\
        Full & Exact $\mid$ HGB+Gre $\mid$ HGB+SVR $\mid$ Avg &25800&$1.735\pm0.001$&$2406.0\pm0.4$&840&30969\\
    \end{tabular}
    \label{tab:MLMC_speedup}
\end{table*}

\section{Results}
\subsection{Experimental setup}\label{sec:setup}

Simulations were implemented in Python 3.8.12 and were run under Windows 10 x64 on a PC equipped with a 4-core Intel Xeon W-2223 CPU (3600 MHz). The optimization problem for the \emph{Avg} model was solved using the python package \emph{quadprog}~\cite{pypi}. The \emph{scikit-learn} package was used for HGBRT and SVR learners in surrogate models~\cite{scikit-learn}. The code is available for download~\cite{MLMC-PMAPS2022}.

For training the surrogate models, 5000 low-margin training days (net generation margin is negative for an hour or more) were generated. Then, the Exact model was used to estimate load curtailment of those days. After that, HGBRT (with the default setting of Scikit-learn 1.0.2) was trained on this dataset, which has 24 input features (hourly net generation margin of a day), to learn $\tilde{f}_{LOL}(\cdot)$. Sampled days that exhibited load curtailment using the Exact model were selected to train the ENS approximation model $\tilde{f}_{ENS}(\cdot)$. To simplify the learning procedure in the light of very large positive margin values, the net generation margin traces were maximised at 1~MW. Next, all features were normalized and the SVR learner was trained to predict the ENS amount. 

To compare MLMC simulations, 6 different architectures based on the aforementioned models in section~\ref{sec:model} were developed (Fig.~\ref{fig:MLMC_correlation}). Estimated risk indices (LOLE and EENS) of simulation results are reported along with the estimated standard error. For each simulation, an exploratory run with $n^{(0)} = 500$ was performed, followed by 50 runs of 500s, where sample sizes were optimized for the EENS risk measure. In all simulations where the Avg model was included, the convolution approach was used to compute $r_0 = E[Y_0]$ without sampling noise. 

\begin{figure}[t!b]
\centerline{\includegraphics[width=8cm]{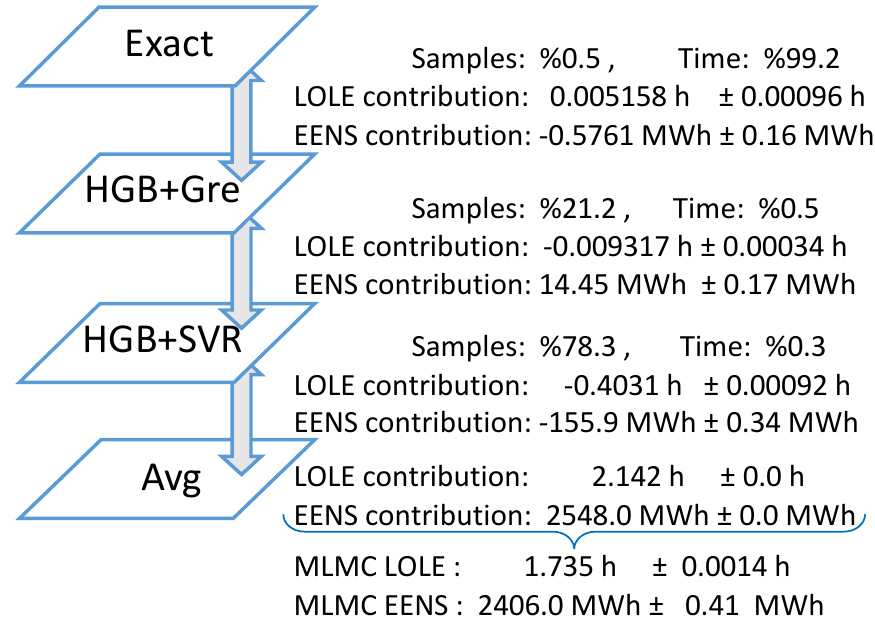}}
\caption{Level pairs and bottom layer contributions for estimating resource adequacy risk indices along with percentage of samples and run time duration for MLMC with Exact, HGB+Gre, HGB+SVR and Avg models. }
\label{fig:MLMC_contribution}
\end{figure}

\subsection{Computational efficiency}

Table \ref{tab:MLMC_speedup} compares the performance of six MLMC architectures with direct MC simulation. All combinations show significant speedups compared to regular MC simulation and the numerical results are consistent with the lack of bias expected from the MLMC method. In all cases, the LOLE speedup is less than the EENS speedup, likely because the discontinuous LOL metric has larger approximation errors than the ENS metric. The \emph{TS20 base/full} results reproduce~\cite{tindemans2020accelerating}, albeit with small changes in observed speedup due to differences in computer architectures and initialisation steps. 

The surrogate-only approach (HGB+SVR) demonstrates the potential for the model to be used without domain-specific models. However, the 3- and 4-layer architectures show that further speedups can be obtained by combining the surrogate models with hand-crafted models, with the 4-layer architecture reaching a speedup of 30,000 for the EENS measure, compared to regular MC sampling.

Figure.~\ref{fig:MLMC_correlation} demonstrates sample correlations between MLMC layers based on LOLE and EENS risk indices. By comparing the two top layers of all MLMC architectures, it becomes obvious that all models are highly correlated to the Exact model which satisfies the first condition of achieving high speedup in simulation. Notably, the HGB+SVR model provides higher LOLE sample correlation (0.998) with Exact model compared to the Avg model (0.989). This is reflected by the higher LOLE speedup for the HGB+SVR model.

It should be noted that the reported speedups are asymptotic values computed from the estimated variance. They do not account for the time required for training the surrogate models. The training time for HGBRT, SVR and generating 5000 low-margin training days were 1, 2, 48 seconds, respectively, which was not included in the simulation results.  

To have a better sense of how MLMC works, layer contributions in the 4-layer MLMC architecture are shown in Fig.~\ref{fig:MLMC_contribution}. The Avg layer has the largest contribution and is the fastest (computed by convolution). Then, each successive level pair refines the estimations of both EENS and LOLE. As we go upward the amount of contribution in estimation decreases while the required time for execution increases.

\subsection{Surrogate model accuracy}

To investigate the effects of training sample size on surrogate models' accuracy and consequently on MLMC performance, this experiment was conducted. 
\begin{enumerate}
    \item Generate 2 sets of 40,000 low-margin days (train and test sets) and compute the EENS-minimising dispatch for each day.
    \item Repeat 100 times for training size=\{500, 1000, 5000\}:
    \begin{enumerate}
        \item Randomly select training samples.
        \item Train the surrogate models (SVR, HGBRT) based on the description given in section~\ref{sec:setup}.
        \item Use test set to compute the Root Mean Square Error (RMSE) of surrogate models.
    \end{enumerate}
    \item Randomly select one of the trained surrogate models for each train size to use in MLMC simulation.
\end{enumerate}
The average RMSE with estimated standard error of 100 runs for each machine learning algorithm and training size is reported in Table~\ref{tab:AI_accuracy}. As expected, better function approximations are obtained with more training samples both for $\tilde{f}_{LOL}$ and $\tilde{f}_{ENS}$.

\begin{table}[tb]
    \caption{Effects of training sample size on surrogate accuracy}
    \renewcommand{\arraystretch}{1.2}
    \begin{tabular}{p{2 cm} p{1.5cm} p{2cm} p{1.5cm}}
    \hline
    Surrogate model & Train size& Average RMSE& RMSE unit\\
    \hline
    SVR&500& $175\pm 10$&MWh/y\\
    SVR&1000&$142\pm 14$&MWh/y\\
    SVR&5000&$97\pm 6$&MWh/y\\
    HGBRT&500&$0.370\pm 0.005$&h/y\\
    HGBRT&1000&$0.269\pm 0.003$&h/y\\
    HGBRT&5000&$0.186\pm 0.001$&h/y\\
    \hline
    \end{tabular}
    \label{tab:AI_accuracy}
\end{table}

To see the effects of surrogate model accuracy and number of training samples on MLMC performance, results of MLMC performance with 3 and 4 levels is presented in Table~\ref{tab:accuracy}. It demonstrates that using a surrogate model, even when trained on a small number of samples, can considerably increase MLMC performance compared to existing approaches. Moreover, further speedups can be obtained by using more training samples.

\begin{table}[tb]

    \caption{Effect of surrogate model accuracy on MLMC performance}
    \renewcommand{\arraystretch}{1.2}
    \begin{tabular}{p{1.6 cm} p{0.8cm} p{0.9cm} p{1 cm} p{1 cm} p{1 cm}}
    \hline
        Estimator & Train size& SVR RMSE (MWh/y)& HGBRT RMSE (h/y)&LOLE Speedup& EENS Speedup\\ \hline
        TS20 full & -   & -    & -   & 125 & 3617 \\
        Hybrid$+$base & 500 & 267 & 0.34 & 349 & 12553 \\
        Hybrid$+$base & 1000 & 153 & 0.23 & 460 & 18723 \\
        Hybrid$+$base & 5000 & 85 & 0.19 & 582 & 20946 \\
        Full & 500 & 267 & 0.34 & 327 & 5248 \\
        Full & 1000 & 153 & 0.23 & 563 & 13546 \\
        Full & 5000 & 85 & 0.19 & 840 &	30969\\
        \hline
    \end{tabular}
   
    \label{tab:accuracy}
\end{table}

\section{Conclusions and future work}
This paper proposes two approaches for using surrogate models with the MLMC method for resource adequacy assessment. One approach is use surrogate models as independent model layers, enabling a `black-box' approach to obtaining speedups with relatively little effort on behalf of the modeller. The other approach  integrates surrogate models with  hand-tuned models for further speedups. The surrogate models can be trained in advance or updated during the simulation run. A resource adequacy case study with storage units was implemented, showing the efficacy of the method on a case that requires time-sequential simulation. The case study shows the superiority of proposed method compared with hand-tuned models in term of speedup. In future work, we will further analyse the impact of training time on optimal speedup, considering the total computational budget. In closing, we emphasise that this approach is not limited to resource adequacy studies; it can be applied to generic risk estimation studies.

\section{Acknowledgments}
The authors thank Zhi Gao for helpful discussions and proof-of-concept experiments.

\printbibliography
\end{document}